\documentclass[editedvolume,numreferences]{crckbked}

\usepackage{times}
\usepackage{graphicx}

\def\largewidth{0.9\textwidth}
\def\figwidth{0.7\textwidth}

\def\smallwidth{0.4\textwidth}

\begin{document}

\tableofcontents

\begin{article}

\begin{opening}
\title{Disordered Wigner crystals}

\author{\surname{T. Giamarchi$^{(1,2)}$}}

\institute{
$^1$ Laboratoire de Physique
des Solides, CNRS-UMR 8502, UPS B\^at. 510, 91405 Orsay France
\\
$^2$ LPTENS CNRS UMR 8549 24, Rue Lhomond 75231 Paris Cedex 05,
France
}

%
%
%
\end{opening}

\section{Introduction}

Disorder effects in quantum electronic systems have led to a
variety of novel phases. Fermionic systems have played a special
role in our understanding of such effects. Indeed for fermions,
the Pauli principle prevent the fermions to be trapped
macroscopically in the minima of the random potential, making the
non interacting case worthwhile to study. Disorder then leads to
the rich physics of Anderson localization. Using both scaling
theories and sophisticated field theoretical techniques, it is now
known that electrons are localized by disorder in one and two
dimensions, whereas a mobility edge exists in three dimensions.
\cite{berezinskii_conductivity_log,abrikosov_rhyzkin,abrahams_loc,%
wegner_localisation,efetov_localisation,efetov_supersym_revue}

The situation becomes much more complicated when one wants to take
into account the electron-electron interaction. Such a question
was crucial for the understanding of doped semiconductors
\cite{thomas_localisation}. In addition recent experiments in two
dimensional electron gas systems have prompted the question of
whether a metal-insulator transition could exist in interacting
systems (see \cite{abrahams_review_mit_2d} and references
therein), stimulating further interest in this problem.

On the theoretical side the question is extremely complicated.
Most of the theoretical approaches used for free electrons either
fail or become much more complicated when interactions are
included which makes it more difficult to obtain unambiguous
answers. Perturbative calculations or renormalization group
calculations can be made for weak interactions. Unfortunately they
scale to strong coupling, which leaves the question of the large
scale/low energy physics still open
\cite{altshuler_aronov,finkelstein_localization_interactions,lee_mit_long}.

In fact, even the pure system is interesting. Indeed for weak
interactions it is reasonable to expect to have a Fermi liquid
behavior, at least for three dimensional systems. Upon increasing
the interaction is was predicted long ago by Wigner that the
electrons would crystallize \cite{wigner_crystal}. Such
crystallization can be induced by other means, decreasing the
density of particles or quenching their kinetic energy by applying
a magnetic field.

This suggests another line of attack for the fermionic disordered
problem: start from the Wigner crystal phase and study the effect
of disorder on such a phase. In the crystalline phase one can
expect the statistics to be less important and thus the problem to
be more tractable. Such a problem falls in the more general
category of disordered elastic systems, which exhibit competition
between elastic forces that like some ordered structure (perfect
crystal for a periodic structure, flat structures for manifolds)
and disorder. Physical systems entering in this category range
from manifold (magnetic domain walls, wetting interfaces etc.),
periodic classical systems (vortex lattice, colloids, magnetic
bubbles, charges spheres, etc.), quantum systems (Luttinger
liquids, stripe phases, charge and spin density waves, Wigner
crystal). There has been an immense activity in these various
domains and many recent progress and it is out of question in
these few pages to cover such a vast topic. I will thus avoid
completely here the topic of classical disordered systems. The
reader who want to know more on that subject and see the links
between the quantum systems discussed here and the classical
problems is referred to the review \cite{giamarchi_cargese_notes},
where this topic is discussed. This review provides references to
other relevant review papers on this vast subject.

For the quantum problems I will confine the discussion to the two
dimensional Wigner crystal. Rather than to try to embark on a
review on the subject, I will focuss on some specific points of
this problem. I will mostly insist on the basic concepts and
present some little additional complements on the question of the
compressibility in such systems. These notes are thus {\it not}
self contained and rely heavily on other published material, both
for contents and references. These few pages can thus be viewed as
an ``appendix'' of the following papers:
\begin{itemize}
\item
The basic concepts of quantum disordered elastic systems are
summarized in the review paper \cite{giamarchi_quantum_revue}. The
reader is thus referred to this paper for the basic physical ideas
in this field and for the technology needed to treat such
problems. In addition this paper contains of course references to
further material. This paper also treats in detail the interesting
case of the one dimensional disordered electron gas. This is a
situation relevant for systems such as quantum wire or nanotubes.
In that case electrons are known to lead to a non fermi liquid
state, the Luttinger liquid, and the disorder effects are
particularly drastic.

\item
The Wigner crystal is examined in details in
\cite{chitra_wigner_hall,chitra_wigner_long}. These papers contain
a detailed discussion of the physical issues, problems and
comparison with the experiments. They describes all the technical
details of the approach that we have used and that I will briefly
discuss in the paper, and give references to other relevant papers
published on the subject.
\end{itemize}

The plan of these notes is the following: In Sec.~\ref{sec:basics}
I will recall the experimental questions as well as the minimal
ingredients underlying the description of a disordered Wigner
crystal. Sec.~\ref{sec:con} discusses the standard approach used
to treat such problems in the past and its comparison with the
recent experiments. Sec.~\ref{sec:quantit} will very briefly
present the results of
\cite{chitra_wigner_hall,chitra_wigner_long} with an emphasis on
the differences between the approach used in Sec.~\ref{sec:con}.
Sec.~\ref{sec:compress} discusses in details the question of the
compressibility and capacitance measurements. Some aspects of the
dynamics are presented in Section~\ref{sec:dynamics}. Conclusions
and perspectives are presented in Section~\ref{sec:conclusion}.

\section{Basic Questions} \label{sec:basics}

\subsection{Experiments}

\begin{figure}
 \centerline{\begin{tabular}{c}
 \includegraphics[width=\smallwidth]{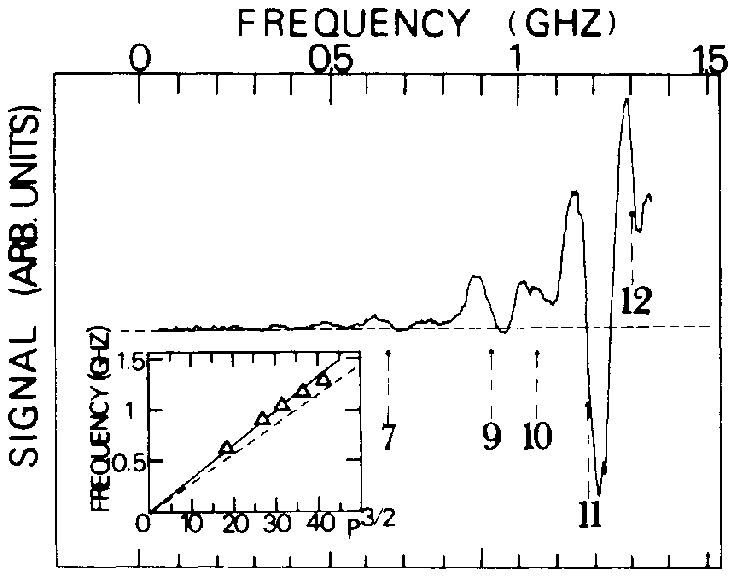}\\
 \includegraphics[width=\smallwidth]{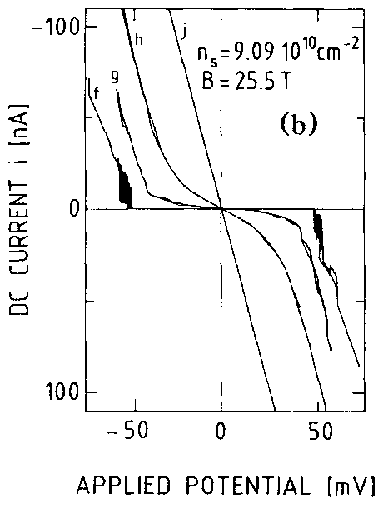}
 \end{tabular}
 \begin{tabular}{c}
 \includegraphics[width=\smallwidth]{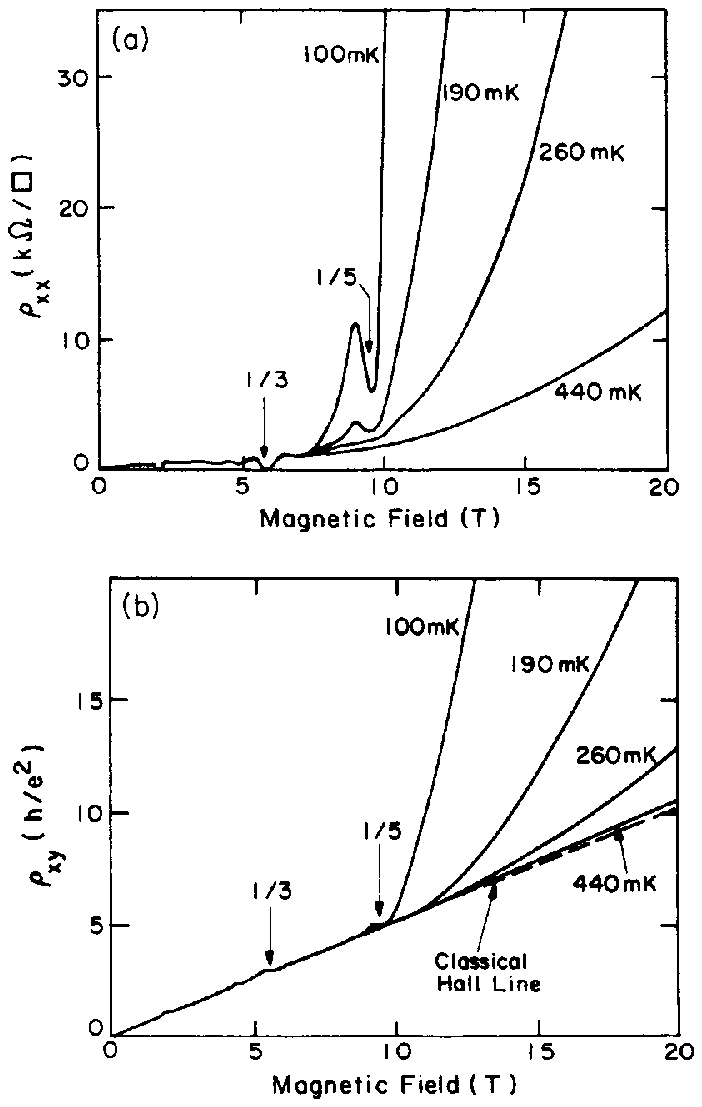}
 \end{tabular}}
\caption{(Top Left) Soundwave absorption by a two dimensional electron gas
(2DEG) under strong magnetic field. The frequencies at which the
sound is absorbed correspond to the eigenmodes of the crystal (see
Sec.~\ref{sec:pure}), and are interpreted as evidence of a Wigner
crystal in this system (from \cite{andrei_wigner_2d}); (Right)
Transport properties of a 2DEG. At strong magnetic field an
insulating phase appears, again suggestive of the formation of a
pinned Wigner crystal (from \cite{willett_wigner_resistivity}).
(Bot. Right) Current vs. voltage characteristics. One clearly sees
a threshold field needed to have conduction. This is again
reminiscent of what one expects of a pinned crystal (From
\cite{williams_wigner_threshold})}
\label{fig:wignerfirst}
\end{figure}

\begin{figure}
\centerline{\includegraphics[width=\figwidth]{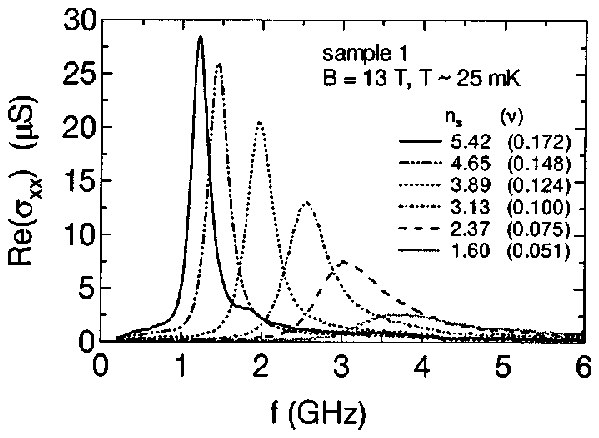}}
\caption{Optical conductivity for various densities for a 2DEG
under strong magnetic field. The peak at a characteristic
frequency (pinning frequency) is again an expected characteristics
of a pinned crystalline structure (see Sec.~\ref{sec:comp}) (from
\cite{li_conductivity_wigner_density}).} \label{fig:optics}
\end{figure}

Although the theoretical concept of a Wigner crystal is easy to
grasp it is much more difficult to check experimentally that such
a phase is realized in nature. Indeed for other crystal states
such as the vortex lattice or the charge density waves, imaging
techniques (decoration, neutrons, X-rays) allow to directly see
the crystal structure. In the case of the Wigner crystal no direct
imaging has been possible so far, although it might become
feasible in a near future. One is thus forced to see whether
indirect measurements (mostly transport, sound absorption) can be
consistently interpreted by assuming a Wigner crystal. The first
evidences for a Wigner crystal are shown in
Fig.~\ref{fig:wignerfirst}.
The optical conductivity, an example of which is shown in
Fig.~\ref{fig:optics}, provides detailed information as we will
discuss it in more details in Sec.~\ref{sec:quantit}.

In all these experiments there is no direct evidence of the
crystal structure. In order to know whether the transport
experiments can be considered as a proof or not of the existence
of the Wigner crystal is it thus specially important to have a
reliable theory that allows to compute the transport properties.
Such a task is far from being trivial given the complexity of the
problem.

\subsection{Elastic description}

\begin{figure}
\centerline{\includegraphics[width=\figwidth]{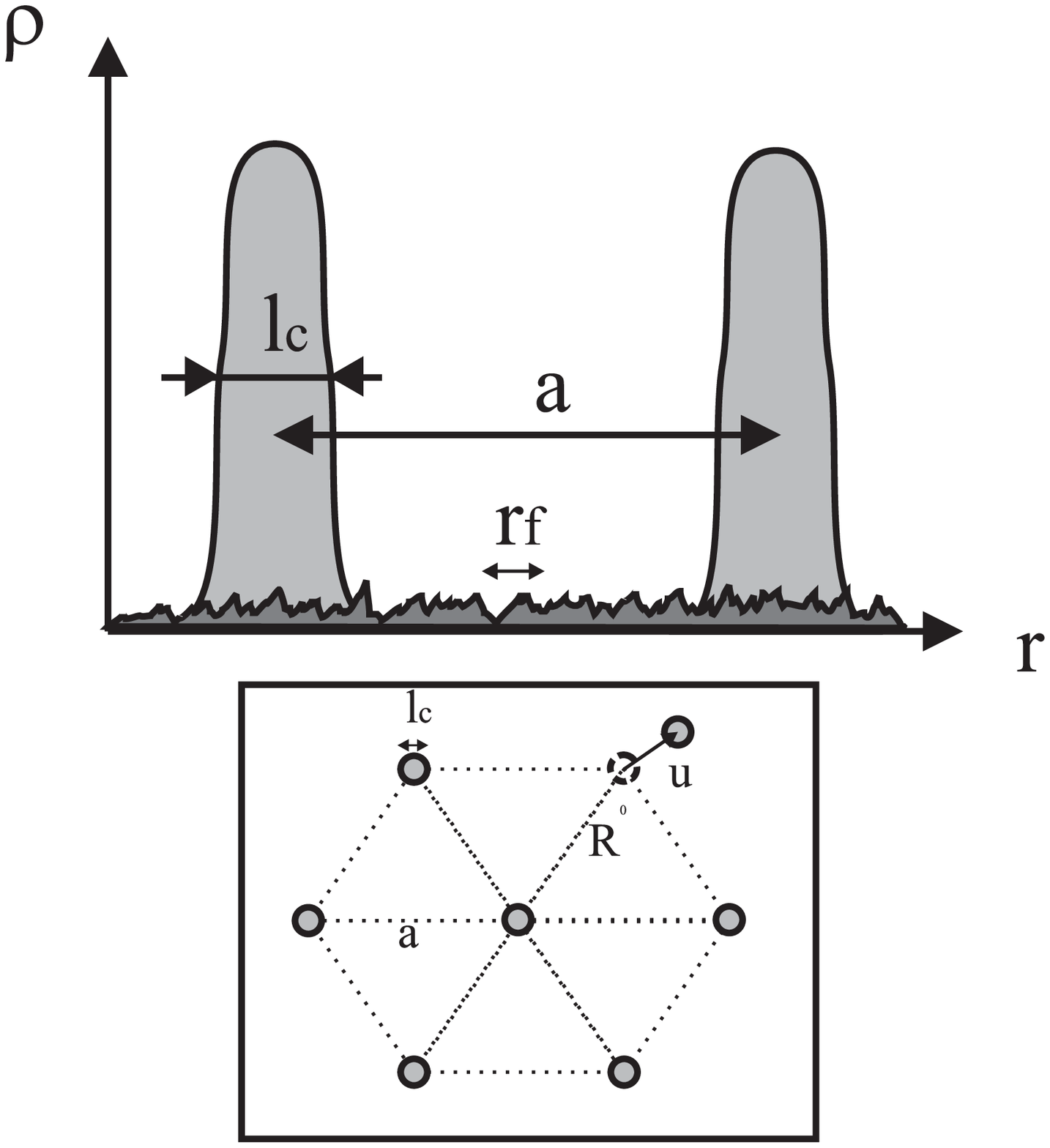}}
\caption{The three length characterizing the Wigner crystal. The
size $l_c$ of the ``particles'' in the crystal (at low temperature
it is essentially given by the extension of the wavefunction
around the equilibrium position, at large temperatures it is
controlled by the thermal fluctuations and is the Lindemann
length), $a$ the lattice spacing is controlled by the density of
particles, and the disorder is correlated over a length $r_f$. The
inset shows  the triangular structure of the Wigner crystal.
Particles are labeled by an equilibrium position ${ R}_i$ and a
displacement ${ u}$.(From \cite{chitra_wigner_long})}
\label{fig:basiclength}
\end{figure}

Starting from the full electronic Hamiltonian (fermions with
interactions and disorder) is a near impossible task. In the
crystal phase some simplifications can be made since the particles
are now discernable by their position. This allows for a minimal
phenomenological model to describe such a crystal: one assumes
that the particle are characterized by an equilibrium position
$R^0_i$ and a displacement $u_i$ relative to this equilibrium
position. In order to define uniquely the displacement one should
not have topological defects such as dislocations in the crystal.
I will come back to this point in Sec.~\ref{sec:con}. From the
original quantum problem one has to define the ``particles'' of
the crystal. If the wavefunction is localized enough then one can
indeed ignore the exchange between the various sites and thus
define ``particles'' that have a size given by the extension of
the wavefunction as indicated in Fig.~\ref{fig:basiclength}.
Of course the density fixes the lattice spacing $a$. These two
lengthscales (size of particle, lattice spacing) are independent
and should be kept. Then one retains for the energy the phonon
modes of the crystal. This leads to the action (see
\cite{chitra_wigner_long} for more details):
\begin{eqnarray}\label{eq:ham}
 S &=& \frac12 \int_{\bf  q} \frac1\beta \sum_{\omega_n}
 u_{L}(q,\omega_n)(\rho_m\omega_n^2 +c_L(q))u_{L}(-q,-\omega_n) \\
 & & + u_{T}(q,\omega_n)(\rho_m\omega_n^2 +c_T(q))u_{T}(-q,-\omega_n) \nonumber \\
 & & + \rho_m \omega_c \omega_n [u_{L}(q,\omega_n)u_{T}(-q,-\omega_n)-
 u_{T}(q,\omega_n)u_{L}(-q,-\omega_n)] \nonumber \\
 & & +  \int d^{2}r \int_0^{\beta } d\tau V({ r})\rho({ r},\tau)
\nonumber
\end{eqnarray}
where we have used the decomposition of the displacements in
longitudinal and transverse modes $\vec{u} = \frac{\vec{q}}{q}
u_L(q) + (\frac{\vec{q}}{q} \wedge \vec{z}) u_T(q)$. $\int_{\bf
q}$ denote the integration over the Brillouin zone $\int_{BZ}
\frac{d^2q}{(2\pi)^2}$, and the $\omega_n$ are the standard
Matsubara frequencies. The third term in (\ref{eq:ham}) comes from
the Lorentz force. $\rho_m \simeq {m \over {\pi a^2}}$ is the mass
density and $\omega_c = e B/m$ the cyclotron frequency.

$C_{L,T}(q)$ are the elastic coefficients for the longitudinal and
transverse modes respectively. These coefficients can be obtained
from an expansion of the coulomb correlation energy of the WC in
terms of the displacements
\cite{bonsall_elastic_wigner,maki_elastic_wigner}. Since the
longitudinal mode describes compressional modes, it is drastically
affected by the coulomb repulsion thus $c_L(q) \propto q$, whereas
the transverse mode describes shear and thus $c_T(q) \propto q^2$
as in elastic media with only short range interactions.

Finally, the last term describes the coupling  to disorder,
modelled here by a random potential $V$. The density of particles
\begin{equation}
\rho(r) = \sum_i \overline{\delta}({ r} - { R}_i - { u}_i)
\end{equation}
where $\overline{\delta}$ is a $\delta$-like function of range
$l_c$ (see Figure~\ref{fig:basiclength}) and ${ u}_i \equiv { u}({
R}_i)$. Since the disorder can vary at a lengthscale $r_f$ {\it a
priori} shorter or comparable to the lattice spacing $a$, the
continuum limit ${ u}_i \to { u}(r)$, valid in the elastic limit
$|{ u}_i - { u}_{i+1}| \ll a$ should be taken with care in the
disorder term \cite{giamarchi_vortex_short,giamarchi_vortex_long}.
This can be done using the decomposition of the density in terms
of its Fourier components
\begin{equation} \label{eq:fourdens}
\rho({ r})\simeq \rho_0 - \rho_0\nabla\cdot { u} + \rho_0 \sum_{{
K} \neq 0} e^{i { K}\cdot({ r} - { u}(r))}
\end{equation}
where $\rho_0$ is the average density and ${ K}$ are the
reciprocal lattice vectors  of the perfect crystal. The finite
range of $\overline{\delta}$ is recovered
\cite{giamarchi_vortex_long} by restricting the sum over $K$ to
momentum of order $K_{\rm max} \sim \pi/l_c$ The disorder is often
assumed gaussian, a limit valid when there are many weak pins
\begin{equation}
\overline{V({ r}) V({ r}')} = \Delta_{r_f}({ r}-{ r}')
\end{equation}
$\Delta_{r_f}$ is a delta-like function of range $r_f$ which is
the characteristic correlation length of the disorder potential
(see Figure~\ref{fig:basiclength}).

These characteristics lengthscales and Hamiltonian define the
minimal model needed to describe a pinned Wigner crystal.

\subsection{Consequences for pure system} \label{sec:pure}

For a pure system the consequences of the quadratic Hamiltonian
(\ref{eq:ham}) are easy to carry out. The eigenmodes of the system
are easy to compute. In the absence of field the longitudinal one
is plasmon like, whereas the transverse one is phonon-like. In the
presence of a large magnetic field the two modes are mixed, giving
the eigenmodes of Table~\ref{tab:modes}.
\begin{table}
\begin{tabular}{c|c|c}
 \hline
 Mode & zero field & High magnetic field \\
 \hline
 $\omega_-(q)$ & $\propto q^2$ (trans.) & $\propto \frac{q^{3/2}}{\omega_c}$ \\
 $\omega_+(q)$ & $\propto q  $ (long.)  & $\sim  \omega_c$ \\
 \hline
\end{tabular}
\caption{$q$ dependence of the eigenmodes in a Wigner crystal in
the absence of magnetic field or for a very strong
field.}\label{tab:modes}
\end{table}
These are the modes that were probed in the sound absorption
experiment shown in Fig.~\ref{fig:wignerfirst}.

For the crystal, the current is simply given by $J = e \rho_0
\partial_t u$, making thus the conductivity very simple to compute
since it is essentially the correlator of the displacements (up to
a factor $\omega$). In the absence of magnetic field the optical
conductivity is a simple $\delta$ function at zero frequency,
which traduces the fact that the crystal slides when submitted to
an external force. In the presence of a finite magnetic field the
electrons describes cyclotron orbits and the peak in conductivity
is pushed to the cyclotron frequency $\omega_c$.

Of course these results are for the pure system only and the
crucial question is to determine how the disorder changes the
above results, in order to make contact with the experiments.

\section{Conventional wisdom} \label{sec:con}

\subsection{Basic ideas and conventional wisdom}

\begin{figure}
\centerline{\includegraphics[width=\figwidth]{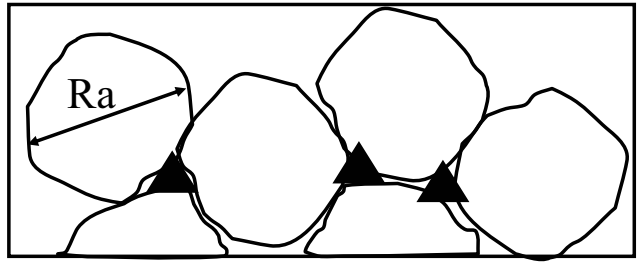}}
\caption{The traditional image of a disordered elastic system. The
system is ``broken'' into crystallites of size $R_a$. The size
corresponds to relative displacements of the order of the lattice
spacing between edges of the ``crystallite''. Topological defects
(e.g. dislocations) are argued to be induced by the disorder at
the same characteristic lengthscale $R_a$. All positional order is
thus lost beyond $R_a$. As we now know (see
Sec.~\ref{sec:quantit}) this physical image is {\it incorrect}.}
\label{fig:conventional}
\end{figure}

In order to know how disorder can modify the above results and
lead to pinning, it is necessary to solve the full problem
(\ref{eq:ham}), an herculean task. People have thus resorted to
approximations. Various such approximation are presented in the
review \cite{giamarchi_quantum_revue}. Based on the various
approximate solutions a conventional wisdom on how a disordered
elastic system should behave has emerged, as shown in
Fig.~\ref{fig:conventional}.
It was believed that because of disorder the crystal is ``broken''
into crystallites whose characteristic size is the pinning length.
Topological defects (dislocations etc.) in the crystals would be
generated at about the same lengthscale. All positional order in
the crystal is thus lost beyond the ``crystallites''. Each
crystallite can thus be seen as pinned practically individually.

\subsection{Comparison with experiments} \label{sec:comp}

\begin{figure}
\centerline{\includegraphics[width=\figwidth]{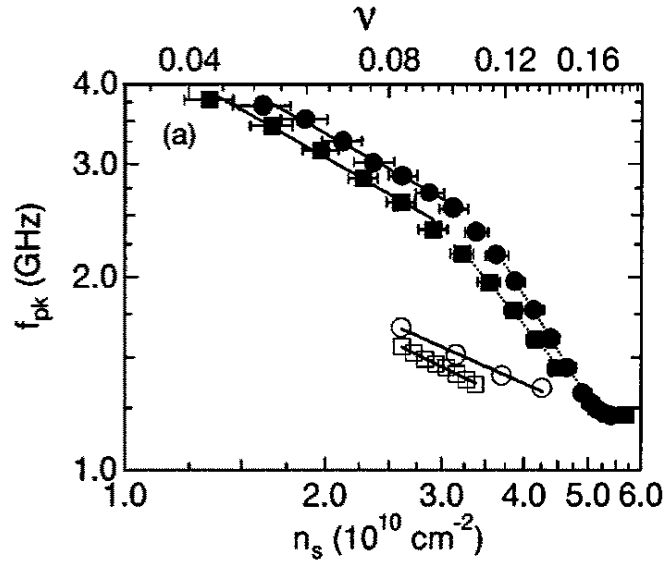}}
\caption{Variation of the pinning frequency with the density for a
2DEG under a strong magnetic field. The pinning frequency
decreases with the density in all systems (different symbols), in
contradiction with the naive calculations based on the physical
image shown in Fig.~\ref{fig:conventional}. Such calculations
would lead to $\omega_p \propto n^{1/2}$. (From
\cite{li_wigner_conductivity_density})} \label{fig:pinfreq}
\end{figure}

This physical image inspired from pioneering theories used for
charge density waves or for the pinning of vortex lattices (see
\cite{giamarchi_quantum_revue} for more details and references)
allows to compute, using some approximations, the optical
conductivity. Essentially the crystallites will be held by a
pinning potential and respond at a given frequency, the pinning
frequency, related to the pinning length, as shown in
Fig.~\ref{fig:pinfreq}. The coupling between the various
crystallites lead to a broadening of the peak, very often assumed
to be lorentzian.

When one compares the experiments, with these theoretical
predictions the agreement is {\it qualitatively} wrong. Let us
show for example the density dependence of the pinning peak (see
Fig.~\ref{fig:pinfreq}).
The predicted density dependence of the pinning frequency would be
from the above mentioned approximations  $\omega_p \propto
n^{1/2}$. This result would be totally opposite to the data which
shows a {\it decrease} of the pinning frequency with the density.
Such important problems when one tries to compare with the data
could cast serious doubts on the interpretation of the insulating
phase in terms of a Wigner crystal and quite naturally other
interpretations for this phase have been proposed
\cite{zhang_hall_insulator}.

Moreover note that in fact the conventional description shown in
Fig.~\ref{fig:conventional} would in fact invalidate the very use
of an elastic approximation such as (\ref{eq:ham}) to compute the
peak in the optical conductivity. Indeed the peak in such approach
would be controlled by the pinning length $R_a$ of the
crystallites. But at that lengthscale topological defects are
argued to occur. Such defects are not taken into account in the
elastic description and could in principle modify the results for
the peak. Strictly speaking the elastic theory could thus only be
applied for frequencies $\omega \gg \omega_p$ (see
Fig.~\ref{fig:disloc}).

\section{Bragg glass and disordered Wigner crystal} \label{sec:quantit}

The situation is in fact much better than one could think based on
the naive approach exposed in Sec.~\ref{sec:con}. In fact the
discrepancy lies in the fact that the theory used to connect the
elastic Hamiltonian (\ref{eq:ham}) to the transport properties and
based on the physical ideas shown in Fig.~\ref{fig:conventional},
is in fact incorrect. Fortunately, it has been possible with
recent ``theoretical technology'' to obtain a quite complete
solution of (\ref{eq:ham}). I will not review here the method or
solution but refer the reader to
\cite{giamarchi_columnar_variat,giamarchi_quantum_revue} for the
general technology and to
\cite{chitra_wigner_hall,chitra_wigner_long} for the Wigner
crystal solution. The agreement with experiments is now quite good
(see \cite{chitra_wigner_long} for a full discussion). In
particular one finds a decrease of the pinning frequency with the
density as $\omega_p \propto n^{-3/2}$, as well as a good magnetic
field dependence of the pinning frequency. This very good
agreement gives a good confirmation that the insulating phase in
the 2DEG under strong magnetic field is indeed a Wigner crystal
collectively pinned by impurities.

Of course despite this good agreement some points still remain
open. Among them the question of the low frequency behavior and
the magnetic field dependence of the width of the peak. I will not
discuss these questions further and refer the reader to
\cite{fogler_pinning_wigner,chitra_wigner_long} for further
discussions of these issues.

I want to insist here on two important physical features which are
apparent in the solution
\cite{chitra_wigner_hall,chitra_wigner_long} and whose physics
deserves to be explained in detail.

The first important point is that the two characteristic
lengthscales (size of particle and lattice spacing) define {\it
two} characteristic lengthscales via the displacement field. The
first one $R_c$, known as the Larkin length, correspond to the
distance for which relative displacements are of the order of the
size of the particle $u(R_c) - u(0) \sim l_c$. The second $R_a$ is
the one for which the relative displacements are of the order of
the lattice spacing $u(R_a) - u(0) \sim a$. Since in the Wigner
crystal $l_c$ and $a$ are quite different $R_c$ and $R_a$
corresponds to quite different lengthscales and have in general
quite different dependence in the various parameters. This is to
be contrasted with charge density waves for which $l_c \sim a$ due
to the nearly sinusoidal density modulation and thus $R_c \sim
R_a$. Thus borrowing directly approximate solutions that have been
developed for this case is dangerous and gives part of the physics
incorrectly. As can be checked from the solution
\cite{chitra_wigner_hall,chitra_wigner_long} the pinning frequency
$\omega_p$ is controlled by the lengthscale $R_c$ and {\it not}
the lengthscale $R_a$. This is physically reassuring since one
knows \cite{larkin_ovchinnikov_pinning} for classical systems that
$R_c$ is the length that controls the threshold force, and thus is
naturally associated with pinning. The distinction is important
since $R_c$ depends on the size of the particle $l_c$. This gives,
for example for the case of strong magnetic field for which $l_c$
is just the cyclotron orbit, additional magnetic field dependence
to the pinning frequency.

\begin{figure}
\centerline{\includegraphics[width=\figwidth]{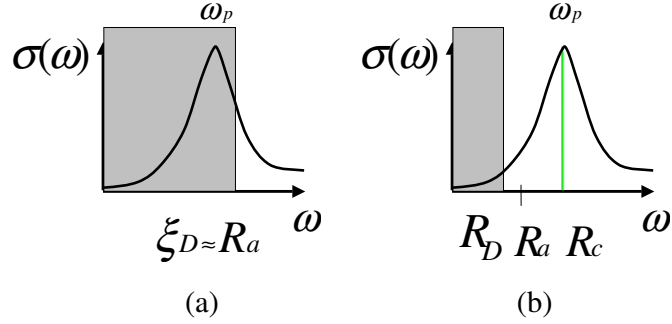}}
\caption{(a) If dislocations  occurred at scale $R_a$ and the
pinning frequency was controlled by $R_a$, as was naively
believed, the elastic theory is incapable of giving any reliable
information on the pinning peak. It would be necessary to include
dislocations from the start. (b) As was discussed in the text,
dislocations occur in fact at $R_D \gg R_a$ and the pinning peak
depends on $R_c \ll R_a$. Thus the pinning peak is given
quantitatively by a purely elastic theory. It is necessary to take
into account topological defects such as dislocations only at much
lower frequencies, and in particular if one wants to obtain
reliable results for the d.c. transport.(from
\cite{chitra_wigner_long})} \label{fig:disloc}
\end{figure}

The second important point concern the possibility to use the
elastic theory. The elastic theory is in fact much more stable to
the presence of topological defects than initially anticipated. In
$d=3$ it is now known that below a certain threshold of disorder
{\it no} topological defects can be induced by the disorder. The
disordered elastic system is in a Bragg glass state
\cite{giamarchi_vortex_long} with a quasi long range positional
order, much more ordered than the image of
Fig.~\ref{fig:conventional} suggests (see e.g.
\cite{giamarchi_cargese_notes} for a discussion and references on
this point). In $d=2$ the situation is marginal, and defects
appear in the ground state, but at distance $R_d$ much larger (for
weak disorder) than the lengthscale $R_a$ and {\it not} at that
lengthscale
\cite{giamarchi_vortex_long,ledoussal_dislocations_2d}.
Dislocations will thus spoil the results of the elastic theory for
the optical conductivity only well below the peak, as shown on
Fig.~\ref{fig:disloc}. This implies that the theory is a {\it
reliable tool} to compute the characteristics of the peak and
above, and thus most of the a.c. transport.

\section{Compressibility} \label{sec:compress}

\subsection{Compressibility in charged systems}

Naively one relates the compressibility to the density-density
correlation function by
\begin{equation} \label{eq:compress}
\kappa = \lim_{q\to 0} \langle \rho(q,\omega_n=0) \rho(-q,\omega_n=0) \rangle
\end{equation}
The case of the disordered system will be discussed in
Sec.~\ref{sec:varcomp}, but let us look first at the pure system.
The compressibility is simply (only the longitudinal mode plays a
role)
\begin{equation}
\kappa(q) = \lim_{q\to 0} \frac{q^2}{c_L(q)}
\end{equation}
If only short range interactions are present in the system the
longitudinal mode is a phonon-like mode $c_L(q) \propto q^2$ and
the one recovers a finite compressibility. On the other hand if
one has long range Coulomb interactions $c_L(q) \propto  |q|$ and
the compressibility becomes zero. This is simply due to the fact
that (\ref{eq:compress}) measures the density response to a change
of chemical potential while {\it keeping the neutralizing
background unchanged}. A charged systems thus does not remain
neutral, hence the infinite compressibility.

One has thus to define the compressibility more precisely. Based
on the correlations a standard substraction procedure consists in
keeping only the ``irreducible'' part of the density-density
correlation function, i.e. define the compressibility as
\begin{equation}
\kappa_{\rm irr} = \frac{\kappa}{1 - V_q \kappa}
\end{equation}
where $V$ is the long range Coulomb potential. However it is
unclear how this procedure is related to the standard way of
measuring the compressibility, i.e. the capacitance measurements
(see below). Many derivations of the compressibility use instead
directly a derivative of the free energy with respect to the
number of particles. The free energy can be computed for a neutral
system for an arbitrary number of particles which solves the
above-mentioned problem. Unfortunately very often the calculation
is only possible in some sort of approximate way such as an
Hartree-Fock approximation. Here again the link with the direct
measurements of the compressibility is not clear. Using such
procedures, so called ``negative'' compressibilities are found for
some range of the interactions, for interacting electrons.
Similarly, experiments measure such negative ``compressibilities''
\cite{eisenstein_hall_compressibility,ilani_compressibility_2DEG}
(see \cite{abrahams_review_mit_2d} for further references and
discussion on this question).

\subsection{Capacitance measurements}

\begin{figure}
\centerline{\includegraphics[width=\figwidth]{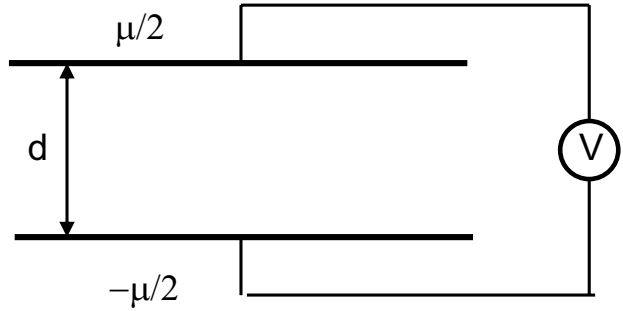}}
\caption{Capacitance measurement, which gives access to the
compressibility of the system. A voltage difference $\mu$ is
applied to a capacitor. Here for simplicity the capacitor is made
of two plates of the 2DEG.} \label{fig:capacitance}
\end{figure}

In order to make the physics of such negative compressibility more
transparent, I will discuss now a very simple way to compute them.
This way is hopefully more physically transparent than the
standard derivations \footnote{This is of course a personal and
probably biased opinion !}, and has the advantage to be easily
extensible to the Wigner crystal in presence of disorder. It is in
fact a direct calculation of the quantity that is actually
measured to determine the ``compressibility'', i.e. the
capacitance of a system made by with the 2DEG
\cite{eisenstein_hall_compressibility}. For simplicity I take here
a capacitor formed of two identical systems, as shown in
Fig.~\ref{fig:capacitance}.
Taking one system and one metallic plate would not change the
results in an essential way. The Hamiltonian of the system is thus
\begin{equation} \label{eq:hamdep}
 H = H_1^0 + H_2^0 + \sum_{(\alpha,\beta)=1,2} \int_{r,r'} \frac12 V(r-r')
 [\rho_\alpha(r) - \rho_0][\rho_\beta(r') - \rho_0]
 + \frac{\mu}2 \int_r [\rho_1(r) - \rho_2(r)]
\end{equation}
If one assumes that the system is neutral in the absence of $\mu$,
then the charge on one plate when a potential $\mu$ is applied is
\begin{equation} \label{eq:congen}
 \langle \rho_1 \rangle = \frac{\mu}2 [ \langle \rho_1 \rho_1 \rangle -
 \langle \rho_1 \rho_2 \rangle ]
\end{equation}
in linear response. (\ref{eq:congen}) give directly the
capacitance $\langle \rho_1 \rangle/\mu$.

As a warmup let us shown that in the RPA approximation
(\ref{eq:congen}) leads back to the standard results for the
compressibility. With (\ref{eq:hamdep}) it is easy to check that
the susceptibilities $\chi_{\alpha\beta} = \langle \rho_\alpha
\rho_\beta \rangle$ are given, in RPA, by
\begin{equation} \label{eq:rpaeq}
 \left(\begin{array}{c} \chi_{11} \\ \chi_{12} \end{array} \right)
 =
 - \left(\begin{array}{cc} \chi^0 V_{11} & \chi^0 V_{12} \\
                         \chi^0 V_{12} & \chi^0 V_{11}
       \end{array} \right)
 \left(\begin{array}{c} \chi_{11} \\ \chi_{12} \end{array} \right)
 +
 \left(\begin{array}{c} \chi^0 \\ 0 \end{array} \right)
\end{equation}
where $\chi^0$ is the bare (i.e. for $H^0$ only) density-density
correlation function in one of the systems. It is easy to solve
(\ref{eq:rpaeq}) to obtain for the (q dependent) capacitance
\begin{equation}
 \chi_{11} - \chi_{12} = \frac{\chi^0}{1 + \chi^0(V_{11} -
 V_{12})}
\end{equation}
The Fourier transform of the Coulomb potentials are given by
\begin{equation}
 V_{11} - V_{12} = \int d^2r e^{i q r} [\frac1r -
 \frac1{\sqrt{r^2+d^2}}] = \frac{(2\pi)(1 - e^{-qd})}{q}
\end{equation}
since the two plates are at a distance $d$. The true capacitance
is the limit $q\to 0$ which leads to
\begin{equation}
C = \frac1{(2\chi^0(q=0))^{-1} + 4 \pi d}
\end{equation}
On thus recovers that the capacitance is the sum of a geometrical
one $C_{\rm geom}$ and one due to the electron gas inside the
plates $C_{\rm el}$
\begin{equation} \label{eq:geom}
\frac1{C} = \frac1{C_{\rm geom}} + \frac1{C_{\rm el}}
\end{equation}
The geometrical one is the standard $1/(4\pi d)$ result. For a
simple electron gas $\chi^0(q=0)^{-1}$ is simply the screening
length $\lambda$ of the electron gas. One thus recovers that the
geometrical distance $d$ between the plates is increased by the
screening length $\lambda$ on each side.

\subsection{Wigner crystal}

One can use the general formula (\ref{eq:congen}) to compute the
capacitance for the Wigner crystal. One substitutes in
(\ref{eq:ham}) the density decomposition (\ref{eq:fourdens}). The
$\nabla u$ terms give directly the contribution of the long range
part of the Coulomb interaction
\begin{equation} \label{eq:lr}
 H_{\rm long-range} = \frac12 \rho_0^2 \sum_q \sum_{\alpha\beta
 =1,2}[V_{\alpha\beta}(q)u_L^\alpha(q)u_\beta(-q)]
\end{equation}
Since $V_{11}(q) \sim 1/q$, (\ref{eq:lr}) gives obviously the part
proportional to $q$ in the elastic coefficients for an isolated
plane. The higher harmonics give the regular part (i.e. the part
proportional to $q^2$ in the elastic coefficients. Such a way to
determine the coefficient is equivalent the calculation of the
coefficients in \cite{bonsall_elastic_wigner}.

Taking a pure system the Hamiltonian becomes (only the
$\omega_n=0$ term of the longitudinal part needs to be computed to
have the compressibility)
\begin{equation}
H =  \left(\begin{array}{c} u_L^1(q) \\ u_L^2(q) \end{array}
\right)
 =
 \left(\begin{array}{cc} c_L^{\rm SR}(q) +  \rho_0^2 q^2 V_{11}(q)& \rho_0^2 q^2 V_{12}(q) \\
                       \rho_0^2 q^2 V_{12}(q)   & c_L^{\rm SR}(q) +  \rho_0^2 q^2 V_{11}(q)
       \end{array} \right)
 \left(\begin{array}{c} u_L^1(-q) \\ u_L^2(-q) \end{array} \right)
\end{equation}
where $c_L^{\rm SR}(q)$ is the ``short range'' part of the elastic
coefficients. Using (\ref{eq:congen}) and the expression of the
density for small $q$ from (\ref{eq:fourdens}) $\rho_L(q) =
-\rho_0 q u_L(q)$ one gets for the capacitance an equation like
(\ref{eq:geom}), where now
\begin{equation} \label{eq:capa}
\frac1{C_{\rm el}} = \lim_{q\to 0} \frac{2 c_L^{\rm
SR}(q)}{\rho_0^2 q^2}
\end{equation}
(again the factor of $2$ comes from the fact that here I took two
identical plates). The electronic one corresponds to the
propagator where {\it only} the short range part of the elastic
coefficients is kept. Using \cite{bonsall_elastic_wigner}
$c_L^{\rm SR}(q) = - \omega_0^2(0.18..)(a q)^2$, where $\omega_0 =
\frac{4\pi e^2}{\sqrt3 m a^3}$, one finds for the Wigner crystal a
``negative'' compressibility. The fact that a system of discrete
charges can lead to such effects has been noted before for
classical Wigner crystals (see e.g.
\cite{nguyen_wigner_biology_houches} and references therein). The
method presented here allows to easily determine this
``compressibility'' and is trivially applicable to the disordered
case.

One should note that if the distance $d$ between the plates is
large compared to the lattice spacing of the Wigner crystal is it
safe to throw away the higher harmonics in the coupling term
$\rho_1 \rho_2$ since they behave as $e^{- K d}$ where $K$ are the
vectors of the reciprocal lattice. The capacitance is then indeed
only given by the properties of a single system as in
(\ref{eq:capa}). This is not true if $d \sim a$ (which can be the
case experimentally) in that case the higher harmonics of the
coupling will also contribute to the capacitance, which then
cannot be trivially related to the intrinsic properties of a
single Wigner crystal. These questions will be addressed in more
details elsewhere.

\subsection{Variational compressibility} \label{sec:varcomp}

Finally let us consider the effects of disorder. The method
discussed in Sec.~\ref{sec:quantit} gives straightforwardly the
density-density correlation function which is simply related to
the displacement-displacement correlation. Since it is a Gaussian
approximation one can easily use the capacitance method shown
above. The capacitance is thus given by (\ref{eq:capa}) but where
one should use the propagator in the presence of disorder. It is
given by (within the variational approximation used
\cite{chitra_wigner_hall,chitra_wigner_long}) by:
\begin{equation} \label{eq:cdis}
 \langle u u \rangle = \frac{1}{\rho_m \omega_n^2 + c_L(q) + \Sigma(1-\delta_{n,0}) + I(\omega_n)}
\end{equation}
where $\Sigma$ and $I(\omega_n)$ are respectively  a constant and
a function related to the disorder verifying $I(0) = 0$.

At $\omega_n=0$ (\ref{eq:cdis}) leads to a compressibility in
identical to the one of the pure system, and thus {\it also}
``negative''. This is the thermodynamics results, but the disorder
can in principle lead to an additional twist. The thermodynamic
compressibility is obviously given by the correlation function in
imaginary time at $\omega_n=0$. For the Wigner crystal this leads
back to the compressibility of the pure system. If however one
makes the analytic continuation $i\omega_n \to \omega + i
\epsilon$, one gets a quite different result. (\ref{eq:cdis})
becomes
\begin{equation}
 \langle u u \rangle = \frac{1}{-\rho_m \omega^2 + c_L(q) + \Sigma + I(\omega)}
\end{equation}
Taking then the limit $\omega \to 0$ first as should be done to
get the compressibility (and not the transport which corresponds
to the opposite limit $q\to 0$ first $\omega \to 0$ after) one has
\begin{equation} \label{eq:glassy}
 \langle u u \rangle = \frac{1}{c_L(q) + \Sigma}
\end{equation}
which leads to a zero compressibility {\it because} of the
presence of the mass $\Sigma$ due to the disorder.  One thus may
expect in an experiment for a disordered Wigner crystal a drastic
change in the behavior of the ``compressibility'' (i.e. the
measured capacitance) of the system as a function of the
frequency. At extremely low frequency, one measures the
thermodynamic compressibility (negative correction for the Wigner
crystal as discussed above). When the frequency increases one
measures (\ref{eq:glassy}), i.e. a zero compressibility. What is
the characteristic frequency separating these two behavior is
still unclear the moment, although it is obviously related to the
coupling to the external environment. This phenomenon seems to be
the equivalent for the Wigner crystal of the Coulomb gap, in usual
interacting electronic systems, where a gap linked to the Fermi
level appears.

\section{Dynamics} \label{sec:dynamics}

We have seen in Sec.~\ref{sec:quantit} that the a.c. transport is
a very efficient way to probe the crystalline nature of the
electronic system. Getting the d.c. transport is a much more
complicated task. For the full quantum system this problem is
still a tough cookie, although it is clear that the same methods
than used for classical system (functional renormalization group
for example) should give good results. One can however gain
considerable intuition on what to expect by looking at the
classical equivalent.

\begin{figure}
\centerline{\includegraphics[width=\largewidth]{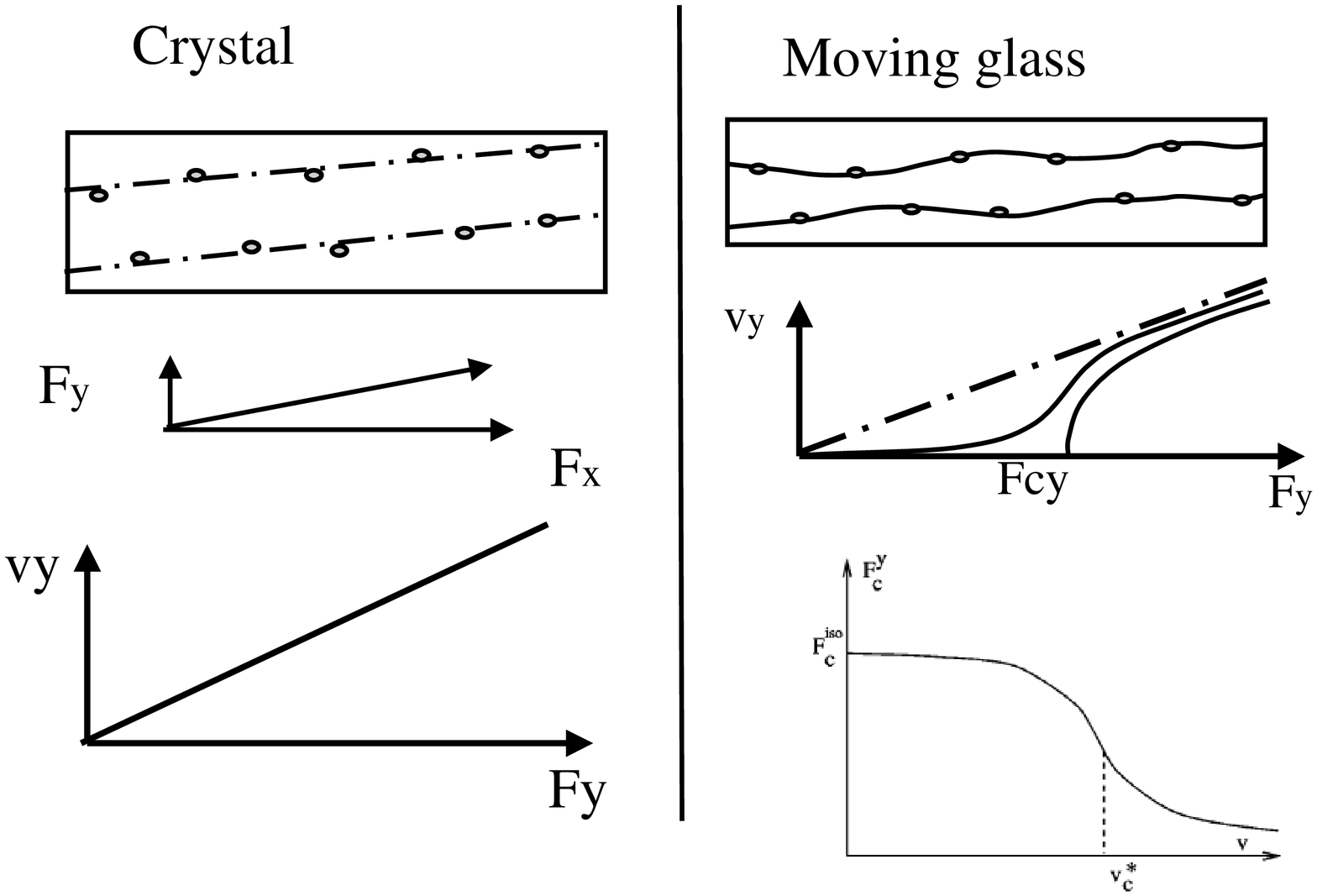}}
\caption{(Left) for a pure crystal the particles move in straight
lines. When the particles move an applied transverse force tilts
the trajectories leading to linear response. (Right) In a
disordered system the moving system is still in a glassy state
(see text). The particle move along rough channels that are the
best compromise between the elasticity and the disorder. Although
the particles themselves move the channels themselves are pinned,
and a transverse pinning force $F_{cy}$ exists. Thus if a
transverse force is applied there is not response (at $T=0$) in
the transverse direction until $F_y
> F_{cy}$. The transverse pinning force decreases with the longitudinal velocity
as shown in the figure above.}
\label{fig:transforce}
\end{figure}

Let us thus consider the case of a two dimensional classical
crystal, submitted to an external force (here due to the electric
field). As was shown in \cite{giamarchi_moving_prl}, periodic
moving periodic systems have a quite specific dynamics. Indeed due
to the existence of the periodicity in the direction {\it
transverse} to the direction of motion, the motion cannot average
completely over the disorder. The moving system is thus submitted
to a random potential, which leads to a channel like motion as
shown in the right hand side of Fig.~\ref{fig:transforce}.

The channels are the best compromise between the elastic energy
and the remaining disorder. Their very existence have an important
consequence if one tries to make the system move in the {\it
transverse} direction. Indeed although the particles do move along
the channels, the channels themselves are pinned. This means that
even above the longitudinal threshold field, if one tries to apply
a force in the transverse direction a transverse critical force
still exists, as shown on Fig.~\ref{fig:transforce}.
The value of this transverse critical force can be computed by
simple scaling arguments \cite{giamarchi_moving_prl} or by more
sophisticated renormalization group techniques
\cite{giamarchi_m2s97_vortex,ledoussal_mglass_long,balents_mglass_long}.

For the two dimensional electronic system, putting a magnetic
field is a simple way of applying a transverse force. If the
lattice is sliding at velocity $v$, it is submitted to a Lorentz
transverse force $F_L = e v B$. The existence of the transverse
critical  force $F_{\rm tr}$ thus implies that the channel
structure should not slide as long as $F_L < F_{\rm tr}$. There
will thus be no hall voltage generated. On the other hand when
$F_L > F_{\rm tr}$ the channel structure should slide and a Hall
voltage exists. The {\it periodicity} of the crystalline structure
thus implies that one needs a {\it finite} longitudinal current
before a Hall voltage exists.

\begin{figure}
\centerline{\includegraphics[width=\smallwidth]{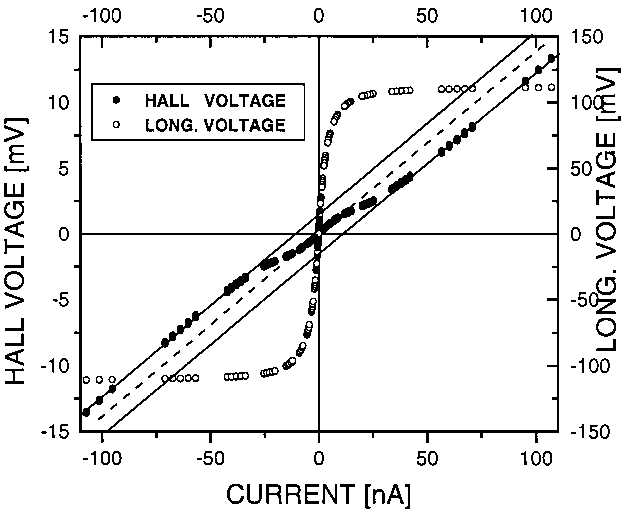}
    \includegraphics[width=\smallwidth]{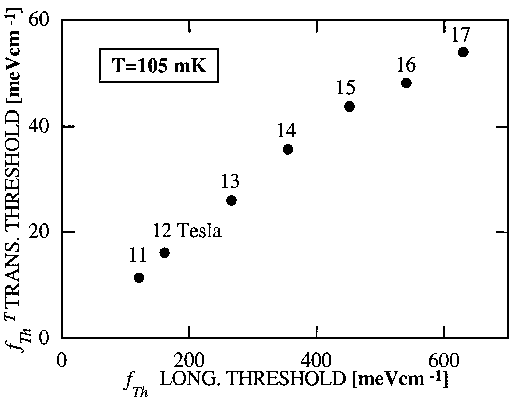}}
\caption{Longitudinal and Hall voltage as a function of the longitudinal
current (left). A nonlinearity is observed in the Hall voltage.
Such a behavior is compatible with the existence of a transverse
threshold. This transverse threshold is plotted vs the
longitudinal one for different magnetic fields (right) (from
\cite{perruchot_wigner_transverse_force})}
\label{fig:tito}
\end{figure}

Such experiment has been performed in the systems under strong
magnetic field \cite{perruchot_wigner_transverse_force}, and the
results are shown in Fig.~\ref{fig:tito}.
A finite longitudinal current is clearly needed to develop a Hall
voltage, in good agreement with the existence of a transverse
threshold. The existence of such an effect is a direct probe of
the crystalline (existence of a transverse periodicity) nature of
the phase.

\section{Conclusions and perpectives} \label{sec:conclusion}

In these short notes I have discussed how the concepts developed
to deal with disordered elastic systems can be fruitfully applied
to interacting electrons. They allow to investigate the effect of
the disorder on a Wigner crystal.

Both the a.c. transport and thermodynamic quantities such as the
compressibility can be reliably computed. There is good agreement
with the predicted d.c. transport and the observed behavior of a
2DEG under a strong magnetic field, making a strong case for a
Wigner crystal phase in such systems. The compressibility is found
to be negative both for a pure Wigner crystal and in the presence
of disorder, and detailed comparison with experiments on that
point would be clearly fruitful. Even if computing the full d.c.
transport is beyond reach at the present, some properties can be
obtained. In particular the periodicity of the crystal should lead
to the existence of a transverse pinning which should entail a
shift in the Hall response, as observed experimentally.

We see that there are thus efficient, if not easy, ways via
transport to check for the presence of a crystalline phase. Most
of the experiments discussed in the preceding sections were for a
2DEG under a strong magnetic field. It would of course be
extremely interesting to use the same techniques to probe the 2DEG
in the absence of the magnetic field, and analyze the experiments
in the line of what was discussed above to check for the existence
of a Wigner crystal in these systems. Among the interesting
possible experiments one can note:
\begin{itemize}
\item
Measurement of the optical conductivity. In particular the
density dependence of the pinning peak can be directly checked
against the theoretical predictions of the pinned Wigner crystal.

\item
If the optical measurements exist, a comparison between the
threshold field in the d.c. transport and the pinning frequency.

\item
The Hall tension vs the longitudinal current (i.e. the measure of
the transverse pinning force)

\item
Although not discussed in these notes, noise measurements are also
a good way to probe the periodic nature of the systems (see e.g.
\cite{togawa_mglass_noise} and references therein).
\end{itemize}

\section{Acknowledgements}

The bulk of the work discussed in these notes results from a
fruitful and enjoyable collaboration with R. Chitra and P. Le
Doussal, both of whom I would like to specially thank. I would
also like to thank E. Abrahams, A. Yacoby and C.M. Varma for many
interesting discussions on the compressibility in charged systems
and F.I.B. Williams for many enlightening discussions on the
Wigner crystal.


\end{article}



\end{document}